# Signal line shapes of Fourier transform cavity-enhanced frequency modulation spectroscopy with optical frequency combs

ALEXANDRA C. JOHANSSON, LUCILE RUTKOWSKI, AMIR KHODABAKHSH, ALEKSANDRA FOLTYNOWICZ*

*Department of Physics, Umeå University, SE-901 87Umeå, Sweden*
*Corresponding author: aleksandra.foltynowicz@umu.se*

We present a thorough analysis of the signal line shapes of Fourier transform-based noise-immune cavity-enhanced optical frequency comb spectroscopy (NICE-OFCS). We discuss the signal dependence on the ratio of the modulation frequency, $f_m$, to the molecular line width, $\Gamma$. We compare a full model of the signals and a simplified absorption-like analytical model that has high accuracy for low $f_m/\Gamma$ ratios and is much faster to compute. We verify the theory experimentally by measuring and fitting NICE-OFCS spectra of $CO_2$ at 1575 nm using a system based on an Er:fiber femtosecond laser and a cavity with a finesse of ~11000.

## 1. INTRODUCTION

Fourier transform spectroscopy with frequency combs [1] offers many advantages over conventional Fourier transform infrared spectroscopy based on incoherent sources. It allows measurement of high resolution spectra with high signal to noise ratios in much shorter acquisition times using a much more compact instrument [2]. Moreover, high absorption sensitivity can be obtained by the use of enhancement cavities [3-6]. The comb can be efficiently coupled into the cavity by matching the repetition rate, $f_{rep}$, to the cavity free spectral range (FSR) and adjusting the offset frequency, $f_{ceo}$, to align the comb lines with the cavity modes. Stable comb-cavity matching can be obtained using two methods – a dither lock, in which the comb or the cavity parameters are swept around the optimum value and the transmitted intensity is averaged over a few modulation periods; or a tight lock, in which $f_{rep}$ and $f_{ceo}$ are locked to the cavity with a high-bandwidth active feedback loop. The dither lock is not compatible with a fast-scanning Fourier transform spectrometer (FTS), in which the interferogram frequency is on the order of a few hundred kHz [4] – the requirement of intensity averaging over a few modulation periods would imply physically impossible dither frequencies in the MHz range. Therefore using a cavity with the FTS requires a tight lock that provides constant power in cavity transmission. Such lock is usually implemented using the two-point Pound-Drever-Hall (PDH) technique, where two error signals are derived at two wavelengths within the comb spectrum to control both degrees of freedom of the comb [5]. The frequency-to-amplitude (FM-AM) noise conversion caused by the residual frequency jitter of the comb with respect to the cavity can then be efficiently removed using two methods: an auto-balancing detector [4], or frequency modulation spectroscopy [7]. In the latter method, called noise-immune cavity-enhanced optical frequency comb spectroscopy (NICE-OFCS) [7], the comb is phase modulated at a frequency equal to (a multiple of) the cavity FSR to produce sidebands to each comb line that are coupled to the cavity as well. Phase sensitive demodulation of the transmitted intensity at the modulation frequency yields a signal that is immune to the FM-AM noise conversion because all comb lines and their sidebands are attenuated and phase shifted by the cavity in the same way. This is similar to what happens in continuous wave noise-immune cavity-enhanced optical heterodyne molecular spectroscopy (NICE-OHMS) [8]. However, the line shapes of molecular signals measured with NICE-OFCS are different than in ordinary frequency modulation spectroscopy (FMS) [9, 10] or NICE-OHMS [11]. The reason is that the NICE-OFCS signal originates from the beating of the comb lines from one FTS arm with the sidebands of the comb lines from the other arm, which are Doppler shifted with respect to each other because of the reflection from the moving mirror. Because of this Doppler shift, which causes different sideband spacing in the two electric fields recombined at the FTS output, the in-phase and out-of-phase NICE-OFCS interferograms consist of a sum of two terms multiplied by slowly varying envelope functions [7, 12]. In particular, there exists an absorption-like term in the out-of-phase interferogram that dominates over all other terms and has a different functional dependence than the dispersion-like signals of FMS and NICE-OHMS. This term has a much weaker dependence on the ratio of the modulation frequency, $f_m$, to the full width at half maximum (FWHM) line width of the absorption line, $\Gamma$, and it is maximized at low modulation frequencies. Moreover, it resides on top of a background that allows normalization and calibration-free absorption measurements. These two properties make the out-of-phase signal the preferred mode of detection in NICE-OFCS.

In our previous work [12] we compared experimental spectra to a simplified model that takes into account only the dominating absorption-like term and neglects the envelopes over the interferograms as well as the comb-cavity offset induced by the two-point PDH lock. While the general agreement was good, the gas concentration received from the multiline fitting had an error of up to 5%, which is too high for precision gas sensing. Moreover, the fit residuals were structured, indicating insufficiencies in the model or distortions in the experimental spectra. In this work we scrutinize the theoretical model of the NICE-OFCS signals to improve the accuracy of concentration retrieval by fitting to experimental data and we discuss the reasons for the discrepancies observed in our previous work. We evaluate to what extent the simplified model, which is computationally less challenging than the full model (i.e. the magnitude of the FFT of the interferogram), can be used to fit to the spectra without sacrificing the accuracy of the concentration retrieval. We determine the optimum range of $f_m/\Gamma$ ratios with respect to the signal size and the applicability of the simplified model. We verify our findings experimentally by measuring NICE-OFCS $CO_2$ spectra using a setup based on an Er:fiber femtosecond laser and a cavity with a finesse of ~11000.

## 2. THEORY

### A. Comb-cavity matching and phase modulation

In NICE-OFCS an optical frequency comb is locked to a cavity and phase modulated at a frequency $f_m$ equal to (a multiple of) the cavity FSR to create sidebands on each comb line. In our previous work we discussed the different possible choices of the FSR/$f_{rep}$ ratio that allow transmitting the comb lines and their sidebands through individual cavity modes [12]. We have shown that one should avoid FSR/$f_{rep}$ ratios for which the sidebands of different comb lines are separated from each other by $f_m$, since the beating between them decreases the stability of the system. To avoid this unwanted interference, empty cavity mode(s) should separate the sidebands belonging to different comb lines. One empty cavity mode separates the sidebands when FSR/$f_{rep} = (2g-1)/4$, where $g$ is a positive integer. For $g$ equal to 1 all comb lines are transmitted, while for higher $g$s the cavity acts as a filter for the comb. The effective repetition rate in cavity transmission is then given by $f_{rep}^T = (2g-1)f_{rep}$. For combs with repetition rates in the hundreds of MHz range, an FSR/$f_{rep}$ ratio of 1/4 yields impractically long cavities. Therefore we found the ratio of 3/4 (i.e. $g$ = 2) to be more convenient, even though it results in the loss of 2/3 of the incident power. In this configuration, shown in Fig. 1, every third comb line is transmitted through every fourth cavity mode, and the effective repetition rate in cavity transmission is equal to $3f_{rep}$. The modulation frequency can be equal to either the FSR or $(4m \pm 1)$ FSR, where $m$ is a positive integer. Figure 1 shows the comb-cavity matching for the two lowest possible modulation frequencies, equal to the FSR in (b), and to 3FSR in (c).

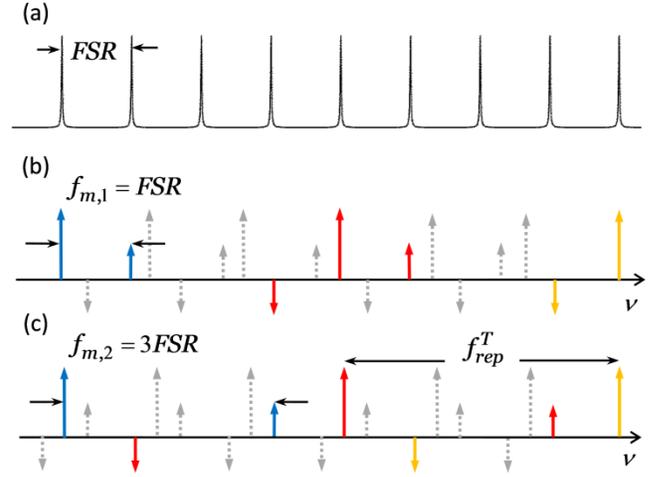

Fig. 1. Comb-cavity matching for FSR/$f_{rep}$ ratio of 3/4 and the two lowest possible modulation frequencies. (a) Cavity modes. (b) Comb spectrum for $f_{m,1}$ = FSR, and (c) $f_{m,2}$ = 3FSR. Solid lines show the transmitted components, each triplet in different color, and dotted gray lines show the reflected components.

### B. Electric field at the cavity output

Provided that the modulation index is small $(\beta \ll 1)$, a sinusoidal phase modulation of an optical frequency comb creates one pair of sidebands for each comb line, separated from the comb lines by $f_m$. The electric field after the cavity is then given by

$$E(\omega,t) = \sum_n \sum_{k=0,\pm 1} \frac{E_n}{2} J_k(\beta) T_{n,k} e^{i[(\omega_n + k\omega_m)t]} + c.c., \quad (1)$$

where $E_n$ and $\omega_n$ are the field amplitude and the angular frequency of the $n$th comb line, $\omega_m = 2\pi f_m$ is the angular modulation frequency, $J_k(\beta)$ is the Bessel function of order $k$, and $T_{n,k}$ is the complex transmission function of the cavity containing the analyte for the $n$th comb line $(k = 0)$ or its sidebands $(k = \pm 1)$ given by [5]

$$T_{n,k} = \frac{te^{-\delta_{n,k} - i\phi_{n,k} - i\varphi_{n,k}/2}}{1 - re^{-2\delta_{n,k} - 2i\phi_{n,k} - i\varphi_{n,k}}}, \quad (2)$$

where in turn $t$ and $r$ are the intensity transmission and reflection coefficients of the cavity mirrors, respectively. The analyte inside the cavity causes a single-pass amplitude attenuation and phase shift of the light given by

$$\delta_{n,k} = \frac{Sc_{rel}pL}{2} \operatorname{Re} \chi_{n,k} \quad (3)$$

and

$$\phi_{n,k} = \frac{Sc_{rel}pL}{2} \operatorname{Im} \chi_{n,k}, \quad (4)$$

respectively. Here $S$ is the line strength (cm$^{-2}$/atm), $c_{rel}$ is the relative concentration of the analyte, $p$ is the gas pressure (atm), $L$ is the cavity length (cm) and $\chi_{n,k}$ is the line shape function (cm). The phase shift picked up during one round-trip inside the cavity is given by

$$\varphi_{n,k} = \frac{2n_r L(\omega_n + k\omega_m)}{c}, \quad (5)$$

where $n_r$ is the index of refraction inside the cavity and $c$ is the speed of light. For comb lines that are exactly on resonance with the cavity modes, this phase shift is equal to a multiple of $2\pi$. However, due to the dispersion in the cavity, which causes cavity FSR to vary with wavelength, only the comb lines at the locking points are exactly on resonance with cavity modes [5, 6]. Away from the locking points the comb lines are transmitted on the slopes of the cavity modes. For these comb lines the round-trip phase shift is given by $\varphi_{n,k} = q2\pi + 2\pi\delta\nu_{n,k}/\text{FSR}$, where $q$ is an integer and $\delta\nu_{n,k}$ is the comb-cavity offset, i.e. the frequency detuning of the comb line/sideband from the center of the cavity mode to which it is locked [5]. This additional phase shift manifests itself as asymmetries in the measured absorption lines.

### C. Electric field at the FTS output

The electric field that is transmitted through the cavity enters the FTS, where it is split by a beamsplitter into two fields that travel in the two interferometer arms. Here we consider an interferometer with both arms scanned simultaneously in opposite directions at a velocity v. The two fields recombined at the output of the FTS are Doppler-shifted in opposite directions because of the reflection from the moving mirrors, and are given by

$$E_\pm = \sum_n \sum_{k=0,\pm 1} \frac{E_n}{4} J_k(\beta) T_{n,k} e^{i\left[(\omega_n + k\omega_m)\left(t \pm \frac{\Delta}{2c}\right)\right]} + c.c., \quad (6)$$

where $\Delta = 4\text{v}t$ is the optical path difference (OPD) between the two arms in the interferometer. Figure 2 shows the frequency domain representation of $E_\pm$ and the beatings at $f_m$ that contribute to the NICE-OFCS signal. The NICE-OFCS interferogram originates from the beating of the comb lines from one interferometer arm with the sidebands of the comb lines from the other arm (and vice versa), indicated by blue dashed arrows. The beating of the comb lines traveling in one interferometer arm with their own sidebands, indicated by red dotted arrows, yields a DC offset after demodulation.

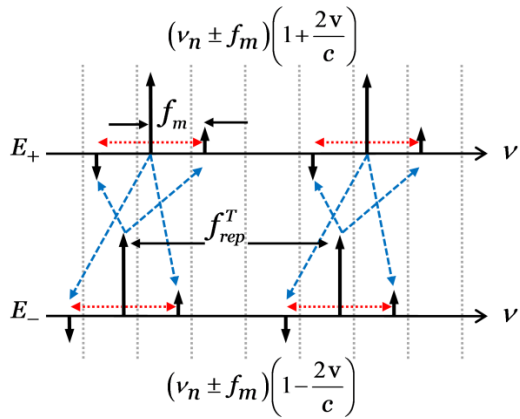

Fig. 2. Frequency domain representation of the two Doppler-shifted electric fields, $E_\pm$, at the FTS output and the possible beatings that contribute to the NICE-OFCS interferogram (blue, dashed) and the DC offset (red, dotted). The vertical dotted grey lines correspond to the optical frequencies that are not Doppler-shifted. Note that the Doppler shift is exaggerated.

### D. NICE-OFCS interferogram

The NICE-OFCS interferogram is obtained by phase-sensitive detection at $f_m$ of the intensity at the output of the FTS, given by $I = c\varepsilon_0 \left\langle (E_+ + E_-)^2 \right\rangle$, and contains one in-phase and one out-of-phase component. To first order approximation, the in-phase component contains information on the dispersion, while the out-of-phase component contains information on the absorption caused by the analyte. The in-phase signal has similar functional form to that of FMS and NICE-OHMS, while the out-of-phase signal is significantly different. As has been discussed in our previous work [12], the out-of-phase component is the preferred mode of detection because it yields a signal that is large for any modulation frequency. In addition it is not zero in the absence of absorption and therefore contains a background that enables normalization and removes the need for calibration. The out-of-phase NICE-OFCS interferogram is given by

$$I_{\omega_m}^{NICE-OFCS} = J_0(\beta) J_1(\beta) \times$$
$$\sum_n I_n \left\{ \cos\left(\omega_m \frac{\Delta}{2c}\right) \cos\left(\omega_n \frac{\Delta}{c}\right) \text{Re}\left(T_{n,0} T_{n,-1}^* - T_{n,0}^* T_{n,1}\right) \right. \quad (7)$$
$$\left. + \sin\left(\omega_m \frac{\Delta}{2c}\right) \sin\left(\omega_n \frac{\Delta}{c}\right) \text{Re}\left(T_{n,0} T_{n,-1}^* + T_{n,0}^* T_{n,1}\right) \right\},$$

where $I_n = c\varepsilon_0 E_n^2$ is the comb line intensity. To first order approximation (i.e. for weak absorption) the first term, multiplied by the cosine envelope, is proportional to $2F(\delta_{n,-1} - \delta_{n,1})/\pi$, where $F = \pi/(1-r)$ is the cavity finesse. It is thus similar to NICE-OHMS absorption signal, and has its maximum when the $f_m/\Gamma$ ratio is close to one, while for lower ratios (i.e. in the undermodulated cases) its magnitude decreases significantly [11]. This is, however, not the case for the second (and dominating) term of the out-of-phase component, multiplied by the sine envelope. To first order approximation this term is proportional to $2 - 2F(\delta_{n,-1} + 2\delta_{n,0} + \delta_{n,1})/\pi$, i.e. it contains a background term, and the attenuations of the comb lines and sidebands add up. Thus, for low modulation frequencies the magnitude of the molecular signal is proportional to $4\delta$, while it decreases to $2\delta$ for higher modulation frequencies, when the contribution from the sidebands and the carrier do not overlap.

The NICE-OFCS interferogram resides on top of a DC offset coming from the beating of the comb lines with their own sidebands, whose out-of-phase component is given by [12]

$$I_{\omega_m}^{OFFSET} = J_0(\beta) J_1(\beta) \times$$
$$\sum_n I_n \cos\left(\omega_m \frac{\Delta}{2c}\right) \text{Re}\left(T_{n,0} T_{n,-1}^* - T_{n,0}^* T_{n,1}\right). \quad (8)$$

This offset is to first order approximation proportional to $2F(\delta_{n,-1} - \delta_{n,1})/\pi$, i.e. it vanishes in the absence of the analyte. Moreover, the amplitude attenuation of the sidebands by the cavity cancels, so the DC offset does not couple in the noise originating from the FM-AM noise conversion by the cavity, which is the key to the noise-immunity of NICE-OFCS.

Figure 3 shows the two terms of the out-of-phase NICE-OFCS interferogram, $\text{Re}(T_{n,0}T_{n,-1}^* - T_{n,0}^*T_{n,1})/2$ in (a) and $\text{Re}(T_{n,0}T_{n,-1}^* + T_{n,0}^*T_{n,1})/2$ in (b), simulated for the $3\nu_1+\nu_3$ $CO_2$ band at 1575 nm for 1000 ppm of $CO_2$ in $N_2$ at a total pressure of 350 Torr in a cavity with a finesse of 11000 and FSR of 187.5 MHz. The spectra are calculated using the complex transmission function from Eq. (2), with each line modeled by a complex Voigt line shape with line parameters from the HITRAN database [13], and a comb-cavity offset set to zero.

The effective repetition rate is assumed to be 750 MHz and $f_{m,1}$ = FSR. The NICE-OFCS interferogram corresponding to these two terms, multiplied by their respective envelopes, is shown in black in (c) in an OPD range of 4 $c/f_{rep}^T$. For comparison, an interferogram for the same $CO_2$ band and for modulation frequency $f_{m,2}$ = 3FSR is shown in (d). In both cases, the intensities of the comb lines, $I_n$, are assumed to follow a Gaussian envelope with a FWHM of ~3 THz. The interferograms consist of three short bursts separated by $c/f_{rep}^T$, whose intensity follows the sine envelopes (solid red and blue curves). At this pressure, the average $CO_2$ FWHM line width is ~2 GHz, which yields an $f_m/\Gamma$ ratio <0.3 for both modulation frequencies. Thus the contribution from the first term [panel (a), multiplied by the cosine envelope, dash-dotted curves] is very small compared to the second term [panel (b), multiplied by the sine envelopes, solid curves] and no burst is visible at zero OPD.

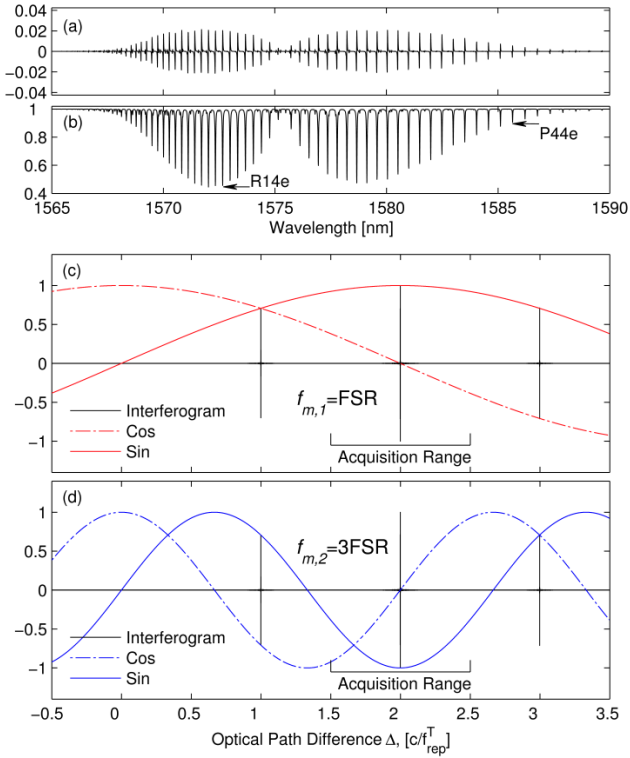

Fig. 3. Simulation of the two terms of the NICE-OFCS interferogram, $\text{Re}(T_{n,0}T_{n,-1}^* - T_{n,0}^*T_{n,1})/2$ in (a) and $\text{Re}(T_{n,0}T_{n,-1}^* + T_{n,0}^*T_{n,1})/2$ in (b), for the $3\nu_1+\nu_3$ $CO_2$ band for 1000 ppm of $CO_2$ in $N_2$ at total pressure of 350 Torr and $f_{m,1}$ = FSR (note the different vertical scales). The two lower panels show NICE-OFCS interferograms based on the simulated spectra (black, normalized to their maximum values) together with the sine and cosine envelopes (solid and dashed curves, respectively) for modulation frequency $f_{m,1}$ = FSR in (c) and $f_{m,2}$ = 3FSR in (d).

### E. NICE-OFCS spectrum

A NICE-OFCS spectrum is obtained by taking the FFT of the interferogram acquired in a symmetric range around any of the bursts at OPD = ±1, 2, 3 $c/f_{rep}^T$. The most intense burst at $2c/f_{rep}^T$ is preferred because it provides highest signal to noise ratio (SNR). The acquisition range around the burst should be limited to $\leq c/f_{rep}^T$, which corresponds to nominal resolution of the NICE-OFCS spectrum of up to $f_{rep}^T$. When the range is equal exactly to $c/f_{rep}^T$ the nominal resolution of the FTS can be exceeded and absorption lines narrower than $f_{rep}^T$ can be measured without distortion by the instrumental line shape function [2]. However, this method requires additional data treatment and stepping of $f_{rep}^T$ to map the entire line shape. Therefore for simplicity we limit our discussion below to absorption lines whose line width is at least 3 times broader than $f_{rep}^T$.

To extract the gas concentration from fits to the measured spectra it is crucial to have a correct model of the signal. The full model of the NICE-OFCS spectrum can be obtained by taking the magnitude of the FFT of the NICE-OFCS interferogram calculated using Eq. (7). However, this procedure is computationally challenging and time consuming as it involves summation over many optical frequencies. In our previous work we used a simplified model of the NICE-OFCS spectrum that is much faster to compute, given by the last dominating factor in the out-of-phase interferogram in Eq. (7), i.e. $\text{Re}(T_{n,0}T_{n,-1}^* + T_{n,0}^*T_{n,1})/2$. This simplified model neglects the contribution from the first term in the out-of-phase interferogram as well as the sine envelope multiplying the second term. We investigate the accuracy of this simplified model by comparing it to the full model for different $f_m/\Gamma$ ratios. For simplicity we focus on one of the strongest absorption lines in the $CO_2$ band simulated above. Figure 4 shows, in black markers, normalized NICE-OFCS spectra of the R14e $CO_2$ line at 1572.66 nm simulated for 1000 ppm of $CO_2$ in $N_2$ using the full model for $f_{m,1}$ = 187.5 MHz [panel (a)] and $f_{m,2}$ = 562.5 MHz [panel (d)] at two different pressures, 350 (solid square markers) and 750 Torr (open circular markers). At these pressures the FWHM molecular line width is 2.1 GHz and 4.5 GHz, i.e. ~3 and ~6 times larger than the nominal resolution given by $f_{rep}^T$ = 750 MHz, and the $f_m/\Gamma$ ratios are 0.09 and 0.04 for $f_{m,1}$ and 0.26 and 0.12 for $f_{m,2}$. The red and blue curves in (a) and (d) (solid for 350 Torr and dotted for 750 Torr) show fits of the simplified model with concentration as the only fitting parameter. The residuals from the fits, shown in (b) and (c) for $f_{m,1}$ and in (e) and (f) for $f_{m,2}$ at the two pressures, respectively, reveal that the simplified model reproduces the full model better when the $f_m/\Gamma$ ratio is smaller. The simplified model matches the full model better for lower modulation frequencies [compare residuals in (b) and (e)], because a more slowly varying envelope multiplying the interferogram affects the signal less. Moreover, for a given modulation frequency, the discrepancy between the models is more pronounced at a lower pressure [i.e. lower line width, compare residuals in (e) and (f)], because the contribution from the first term in the out-of-phase interferogram [Eq. (7)] is larger for higher $f_m/\Gamma$ ratios. The concentrations received from the fits of the simplified model for $f_{m,1}$ are 1000 ppm at 350 Torr and 1000 ppm at 750 Torr. The corresponding values for $f_{m,2}$ are 1005 and 1001 ppm, respectively. Hence, even for the highest $f_m/\Gamma$ ratio shown here, i.e. 0.26, the error on the concentration is only 0.5%.

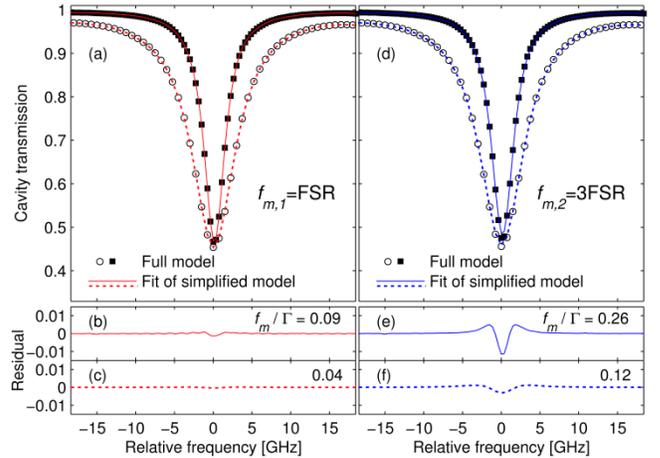

Fig. 4. NICE-OFCS spectra of the R14e $CO_2$ line at 1572.66 nm simulated using the full model at 350 Torr (FWHM = 2.1 GHz, zero padded by a factor of 2, solid square markers) and 750 Torr (FWHM = 4.5 GHz, open circular markers) for two modulation frequencies, (a) $f_{m,1}$ = FSR and (d) $f_{m,2}$ = 3FSR. The red and blue curves show fits of the simplified model, solid curve for 350 Torr and dotted for 750 Torr. Residuals of the fits at the two line widths are shown in (b) and (c) for $f_{m,1}$, and in (e) and (f) for $f_{m,2}$, and the corresponding $f_m/\Gamma$ is indicated in each panel.

Figure 5 shows the peak-to-peak value of the residual of the fit of the simplified model to the full model as a function of $f_m/\Gamma$ for two $CO_2$ lines with different absorption [indicated by arrows in Fig. 3(b)]: the P44e line with ~10% absorption (black square markers) and the R14e line with ~50% absorption (red circular markers), normalized to the line intensities. The four cases shown in Fig. 4 are indicated with black circles. The figure shows that the discrepancy between the full and simplified model increases with the $f_m/\Gamma$ ratio and it is higher for the line with lower absorption. However, at low modulation frequencies, below 0.1 $\Gamma$ for the line with 10% absorption, and below 0.15 $\Gamma$ for the line with 50% absorption, the peak-to-peak value of the residual is below 1% of the line intensity, which means the discrepancy would not be visible for lines with a SNR of 100. Therefore there is an advantage to use low $f_m/\Gamma$ ratios because the amplitude of the signal is maximized and the simplified model can be used for fitting to the experimental data without loss of accuracy. This implies that the modulation frequency should be chosen lowest possible, i.e. equal to the cavity FSR, and the optimum range of pressures is then found from Fig. 5 and the SNR in the spectrum.

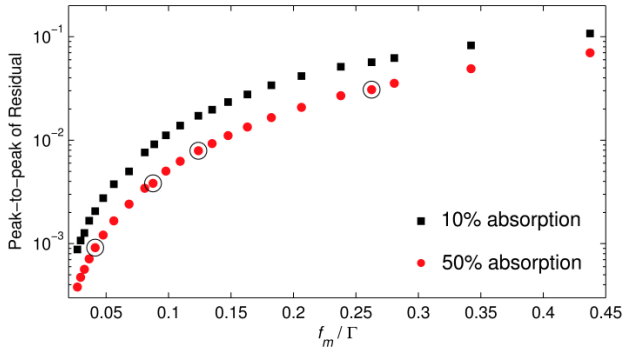

Fig. 5. Peak-to-peak value of the residual of fitting of the simplified model spectra to the full model spectra as a function of the $f_m/\Gamma$ ratio for for the P44e line with 10% absorption (black square markers), and for the R14e $CO_2$ line with 50% absorption (red circular markers), normalized to the line intensity. The cases shown in Fig. 4 are indicated with black circles.

## 3. EXPERIMENTAL SETUP AND PROCEDURES

The experimental setup, depicted in Fig. 6, follows in large the setup described in [12]. The comb source is an Er:fiber laser emitting in the 1.5-1.6 μm wavelength range with a repetition rate of 250 MHz. The comb is locked to an enhancement cavity, using the two-point Pound-Drever-Hall (PDH) locking technique [5]. The cavity is made of two highly reflective concave mirrors mounted on a stainless steel spacer tube with a length of 80 cm. The cavity FSR is 187.5 MHz, so every third comb line is transmitted through every fourth cavity mode, yielding effective repetition rate, $f_{rep}^T$, of 750 MHz. The cavity finesse, measured with 1% accuracy by cavity ring down, is ~11000 around 1575 nm, which corresponds to a cavity mode FWHM of ~17 kHz. The cavity is connected to a gas flow system where the available gases are 1000(2) ppm of $CO_2$ in $N_2$ and pure $N_2$. These gases are used to measure the NICE-OFCS signal and background, respectively.

A fiber-coupled EOM is used to generate the sidebands for PDH locking and NICE-OFCS detection. We use two different NICE-OFCS modulation frequencies ($f_{m,1} = f_{rep}^T/4$ = FSR = 187.5 MHz and $f_{m,2} = 3f_{rep}^T/4$ = 3FSR = 562.5 MHz), both with a modulation index of 0.33. The modulation frequency is generated by a direct digital synthesizer (DDS) referenced to the fifth harmonic of the repetition rate. The DDS output is set to 3/4 $f_{rep}$ or 9/4 $f_{rep}$, which ensures that the sidebands are passively locked to their cavity modes even if the cavity length drifts [12].

The light transmitted through the cavity is sent into a home-built fast-scanning FTS. For both modulation frequencies we acquire the interferogram centered at OPD = 80 cm (i.e. $2c/f_{rep}^T$) in a range of ±20 cm (i.e. $c/f_{rep}^T$), which yields a nominal resolution of 750 MHz, corresponding to the spacing between the transmitted comb lines. The scan over the interferogram takes 0.5 s, and the total acquisition time is 0.9 s caused by the dead time when the retro reflector changes direction. The comb interferogram is measured with a 1 GHz bandwidth InGaAs detector, bandpass filtered, amplified, and demodulated at the modulation frequency using phase-sensitive detection to yield the NICE-OFCS interferogram. The detection phase is adjusted by maximizing the interferogram amplitude using a phase-shifter placed between the DDS output and the local oscillator input of the mixer. The OPD is calibrated using a stabilized HeNe laser, whose beam is propagating parallel to the comb beam. The comb and HeNe interferograms are recorded with a 2-channel data acquisition card at the rate of 5 MSample/s and 20 bit resolution. The NICE-OFCS interferogram is resampled at the zero-crossings and extrema of the HeNe interferogram. The NICE-OFCS spectrum is obtained by taking the magnitude of the FFT of the NICE-OFCS interferogram.

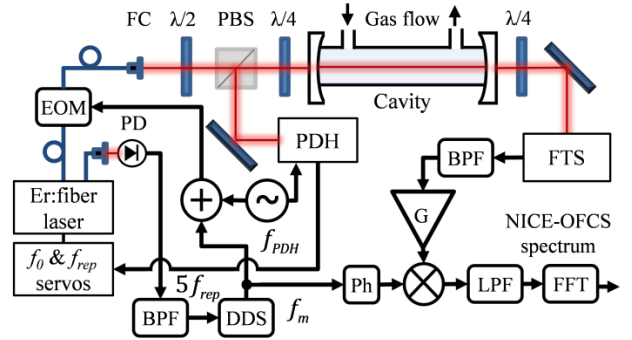

Fig. 6. Experimental setup. EOM – electro-optic modulator, FC – fiber collimator, λ/2 – half-wave plate, PBS – polarizing beam splitter, λ/4 – quarter-wave plate, FTS – Fourier transform spectrometer, PD – photodetector, PDH – Pound-Drever-Hall error signal generation, BPF – band-pass filter, DDS – direct digital synthesizer, G – amplifier, LPF – low-pass filter, Ph – phase shifter, FFT – fast Fourier transform.

## 4. RESULTS

In our previous work we used the maximum power available on the detector in the FTS (300 μW) in order to maximize the SNR [7, 12]. However, when scrutinizing the signal line shapes we observed saturation effects that introduced distortion and caused a 5-10% error on retrieved concentration. The detector saturation reduces the burst intensity compared to the wings of the interferogram, which manifests itself as higher intensities of the absorption lines after the FFT, which in turn yields an elevated gas concentration from the fit and a structure in the residual. To investigate the effect of saturation we measured the NICE-OFCS signals at different power levels and studied the change of received gas concentration and the remaining residual from fits to the experimental data. We found the highest optical power on the detector for which we did not notice the effects of saturation to be 100 μW. Therefore all spectra are measured with this power level on the detector and averaged 100 times to reach SNR of 500.

### A. Line shape verification

To verify the results of the simulations shown in section 2E, we measured NICE-OFCS spectra of $CO_2$ for two modulation frequencies, $f_{m,1}$ (187.5 MHz) and $f_{m,2}$ (562.5 MHz), at two pressures, 350 and 750 Torr, that correspond to the $f_m/\Gamma$ ratios shown in Fig. 4. Figure 7 shows the experimental NICE-OFCS spectra of the R14e $CO_2$ line together with fits of the simplified model, calculated the same way as in section 2E. The line parameters (center frequency, line strength and pressure broadening) are fixed to the values from the HITRAN database. The intensity transmission and reflection coefficients of the cavity mirrors are calculated from the experimentally determined value of the finesse at this wavelength (~10600). The fitting parameters are $CO_2$ concentration and the comb-cavity offset, which is found to be ~0.16 kHz. The experimental data (after baseline correction) is shown in black markers and the corresponding fitted $CO_2$ spectra are shown in red for $f_{m,1}$ in (a) and in blue for $f_{m,2}$ in (d), for 350 Torr in solid curves and 750 Torr in dotted curves. The residuals from the fits at the two pressures are shown in (b) and (c) for $f_{m,1}$, and in (e) and (f) for $f_{m,2}$, respectively. Except for some remaining etalons, the residuals have magnitude and structure similar to those shown in Fig. 4. The simplified model agrees better with the spectra measured with the lower modulation frequency, 187.5 MHz, than with 562.5 MHz, which is in agreement with the simulations. For the higher modulation frequency, the peak-to-peak value of the residual is higher at the lower pressure, which is also in agreement with the predictions from simulations. The SNR is 500 in the spectra measured with $f_{m,1}$ [panel (a)], and slightly lower for $f_{m,2}$ [panel (d)]. In fact, the noise in (b) and (c) does not allow observing the structure expected in the residual, which is justified by the predictions of the peak-to-peak value of the residual between the simplified and full model at these conditions shown in Fig. 5. Hence, with this SNR the simplified model is fully sufficient to model the experimental spectra. The concentrations received from the fits with $f_{m,1}$ are 1018(1) ppm and 1009(1) ppm for 350 and 750 Torr, respectively, and the corresponding values for $f_{m,2}$ are 1026(2) ppm and 1014(1) ppm, respectively. The difference between the four concentrations and the discrepancy from the expected value [1000(2) ppm] is larger than the combined error from the fits and from the use of the simplified model (<0.5%). The variation is caused by the 1% uncertainty of the gas pressure controller and the need to refill the cavity for each measurement, and the higher experimental value is caused by the 1% uncertainties on the gas pressure and cavity finesse measurements.

### B. Multiline fit of NICE-OFCS spectrum

The fits to a single $CO_2$ line shown above verify the validity of the presented model as well as the comparison between the full and simplified models. In order to take advantage of the multiline fitting that offers improved precision in the gas concentration retrieval we now make a fit to the entire $CO_2$ band. Figure 8(c) shows in black the normalized NICE-OFCS spectrum of the $3\nu_1+\nu_3$ $CO_2$ band at 350 Torr measured with the lower modulation frequency of 187.5 MHz. The red curve, inverted for clarity, shows a fit of the simplified model spectrum, where each line is calculated using a Voigt profile with parameters from the HITRAN database, and the experimentally determined values of cavity finesse and comb-cavity offset. Cavity finesse, measured by cavity ring-down, is shown with black markers in (a) together with a third-order polynomial fit (red). The comb-cavity offset, shown with black markers in (b), is found from fits to the individual $CO_2$ lines, similar to that shown in Fig. 7, and the red curve is a third-order polynomial fit. These polynomial fits for finesse and comb-cavity offset are used as input parameters in the model of the entire $CO_2$ spectrum. The residual of the fit, shown in (d), demonstrates that the general agreement between the measurement and the model is very good over the entire spectrum. The small remaining residual comes mostly from errors in the determination of the comb-cavity offset. The concentration found from the fit is 1017.8(2) ppm, in agreement with the value obtained from a single line fit, but with better precision.

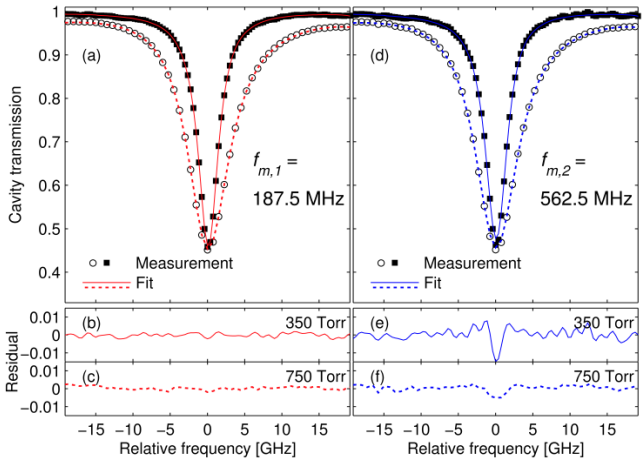

Fig. 7. Experimental NICE-OFCS spectra of the R14e $CO_2$ line (black markers, 100 averages) at a pressure of 350 Torr (solid squares, zero padded by a factor 2) and 750 Torr (open circles) together with fits of the simplified model at the two modulation frequencies, $f_{m,1}$ = 187.5 MHz in (a) (red) and $f_{m,2}$ = 562.5 MHz. in (d) (blue). Residuals of the fits are shown in (b) and (c) for $f_{m,1}$ and in (e) and (f) for $f_{m,2}$, at 350 and 750 Torr, respectively.

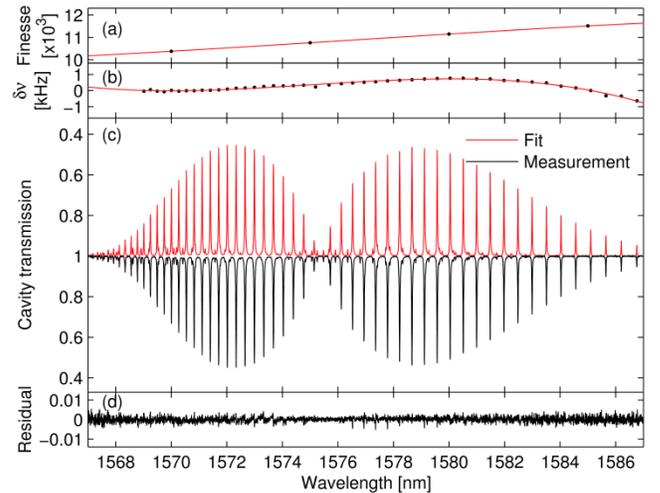

Fig. 8. NICE-OFCS spectrum of the $3\nu_1+\nu_3$ $CO_2$ band at 350 Torr for the lower modulation frequency (187.5 MHz). (a) Cavity finesse measured by cavity ring-down (black markers) together with a third-order polynomial fit (red). (b) Comb cavity offset, $\delta\nu$, found from line by line fitting (black markers) together with a third-order polynomial fit (red). (c) Normalized NICE-OFCS spectrum (black, 100 averages) of 1000 ppm $CO_2$ in $N_2$ at total pressure of 350 Torr, together with a fit of the simplified model (red, inverted). (d) Residual from the fit.

## 5. SUMMARY AND CONCLUSIONS

The NICE-OFCS signal line shape is different from those in continuous wave FMS and NICE-OHMS because it originates from the beating of two phase-modulated fields at the output of the FTS that are Doppler shifted with respect to each other. The NICE-OFCS signal has an absorption-like shape and it is maximized at low modulation frequencies. Moreover, it resides on top of a background that enables normalization and makes the NICE-OFCS technique calibration free. It should be noted that if the phase modulation was performed after the FTS, the resulting signal line shape would resemble that obtained in FMS [14]. However, in NICE-OFCS the phase modulation must be applied before the FTS, because the Doppler shift prevents coupling of the light to a cavity after the FTS.

The full model of the NICE-OFCS spectra involves an FFT of a simulated interferogram, so it is time-consuming to calculate and does not allow fast fitting to the data. Therefore we made a quantitative study of the accuracy of the simplified analytical model, which takes into account only the dominating absorption-like term and neglects the sine envelope multiplying the interferogram. We found that the accuracy of this simplified model depends on the $f_m/\Gamma$ ratio and that it is best for low $f_m/\Gamma$ ratios. In particular, for a line with 50% absorption the accuracy of the simplified model is better than 1% for $f_m/\Gamma$ ratios below 0.15, given by the peak-to-peak value of the residual between the two models. Such discrepancy between the models has little effect on the gas concentration retrieved from the fit. For the same line with 50% absorption it introduces an error smaller than 0.5%, which is often below other systematic uncertainties in the experiment. In general, the limitations of the simplified model will be apparent only when the relative noise in the measured signal is lower than the discrepancy between the models. Thus the simplified model can be used without loss of accuracy when the relative noise is higher than the residual between the two models, shown in Fig. 5. To summarize, the modulation frequency for a NICE-OFCS measurements should be chosen lowest possible, i.e. equal to the cavity FSR, in order to maximize the signal. The range of pressures in which the simplified model can be used for fast fitting to the spectra without loss of accuracy is then determined by the experimental SNR.

We found that the structures visible in the residuals of the fits shown in our previous work [12] had two origins: distortion caused by saturation of the FTS detector, and inadequacies of the theoretical model used for fitting. Reducing the power level on the detector allowed measuring undistorted NICE-OFCS spectra of the $3\nu_1+\nu_3$ $CO_2$ band with SNR of 500 in 90 s. Fitting of the simplified model including the proper value of the comb-cavity offset yielded nearly structureless residuals for the spectra measured with the lower modulation frequency. Thus the accuracy of the retrieved concentration is not limited by the theoretical model used for fitting, but instead is given by systematic experimental errors (mainly the inaccuracy of the pressure controller).

**Funding Information.** Swedish Research Council (621-2012-3650); Swedish Foundation for Strategic Research (ICA12-0031); Stiftelsen J C Kempes Minnes Stipendiefond (FS 2.1.6-217-16).

## REFERENCES

1. Mandon, J., G. Guelachvili, and N. Picque, "Fourier transform spectroscopy with a laser frequency comb," Nat. Photonics **3**, 99-102 (2009).
2. Maslowski, P., K.F. Lee, A.C. Johansson, A. Khodabakhsh, G. Kowzan, L. Rutkowski, A.A. Mills, C. Mohr, J. Jiang, M.E. Fermann, and A. Foltynowicz, "Surpassing the path-limited resolution of Fourier-transform spectrometry with frequency combs," Phys. Rev. A **93**, 021802 (2016).
3. Kassi, S., K. Didriche, C. Lauzin, X.D.D. Vaernewijckb, A. Rizopoulos, and M. Herman, "Demonstration of cavity enhanced FTIR spectroscopy using a femtosecond laser absorption source," Spectroc. Acta A **75**, 142-145 (2010).
4. Foltynowicz, A., T. Ban, P. Maslowski, F. Adler, and J. Ye, "Quantum-noise-limited optical frequency comb spectroscopy," Phys. Rev. Lett. **107**, 233002 (2011).
5. Foltynowicz, A., P. Maslowski, A.J. Fleisher, B.J. Bjork, and J. Ye, "Cavity-enhanced optical frequency comb spectroscopy in the mid-infrared - application to trace detection of hydrogen peroxide," Appl. Phys. B **110**, 163-175 (2013).
6. Maslowski, P., K.C. Cossel, A. Foltynowicz, and J. Ye, Cavity-enhanced direct frequency comb spectroscopy, in *Cavity-Enhanced Spectroscopy and Sensing*, vol. 179 of Springer Series in Optical Science, H.P. Loock and G. Gagliardi, eds. (Springer, 2013). pp. 271-321.
7. Khodabakhsh, A., C. Abd Alrahman, and A. Foltynowicz, "Noise-immune cavity-enhanced optical frequency comb spectroscopy," Opt. Lett. **39**, 5034-5037 (2014).
8. Ye, J., L.S. Ma, and J.L. Hall, "Ultrasensitive detections in atomic and molecular physics: demonstration in molecular overtone spectroscopy," J. Opt. Soc. Am. B **15**, 6-15 (1998).
9. Bjorklund, G.C., "Frequency-modulation spectroscopy: a new method for measuring weak absorptions and dispersions," Opt. Lett. **5**, 15 (1980).
10. Bjorklund, G.C., M.D. Levenson, W. Lenth, and C. Ortiz, "Frequency modulation (FM) spectroscopy," Appl. Phys. B **32**, 145-152 (1983).
11. Ehlers, P., I. Silander, and O. Axner, "Doppler-broadened NICE-OHMS - optimum modulation and demodulation conditions, cavity length, and modulation order," J. Opt. Soc. Am. B **31**, 2051-2060 (2014).
12. Khodabakhsh, A., A.C. Johansson, and A. Foltynowicz, "Noise-immune cavity-enhanced optical frequency comb spectroscopy: A sensitive technique for high-resolution broadband molecular detection," Appl. Phys. B **119**, 87-96 (2015).
13. Rothman, L.S., I.E. Gordon, Y. Babikov, A. Barbe, D.C. Benner, P.F. Bernath, M. Birk, L. Bizzocchi, V. Boudon, L.R. Brown, A. Campargue, K. Chance, E.A. Cohen, L.H. Coudert, V.M. Devi, B.J. Drouin, A. Fayt, J.M. Flaud, R.R. Gamache, J.J. Harrison, J.M. Hartmann, C. Hill, J.T. Hodges, D. Jacquemart, A. Jolly, J. Lamouroux, R.J. Le Roy, G. Li, D.A. Long, O.M. Lyulin, C.J. Mackie, S.T. Massie, S. Mikhailenko, H.S.P. Mueller, O.V. Naumenko, A.V. Nikitin, J. Orphal, V. Perevalov, A. Perrin, E.R. Polovtseva, C. Richard, M.A.H. Smith, E. Starikova, K. Sung, S. Tashkun, J. Tennyson, G.C. Toon, V.G. Tyuterev, and G. Wagner, "The HITRAN2012 molecular spectroscopic database," J. Quant. Spectrosc. Radiat. Transf. **130**, 4-50 (2013).
14. Mandon, J., G. Guelachvili, and N. Picqué, "Frequency-modulation Fourier transform spectroscopy: a broadband method for measuring weak absorptions and dispersions," Opt. Lett. **32**, 2206-2208 (2007).